\def \myStyle {\styleJOSAB}
\def \styleJOSAB {\styleJOSAB}
\def \stylePRA {\stylePRA}

\ifx \myStyle \stylePRA
\documentclass[aps,pra,reprint,groupedaddress,showpacs]{revtex4-1}
\fi

\ifx \myStyle \styleJOSAB
\documentclass[letterpaper,12pt]{article}   
\usepackage{osajnl2} 
\fi

\usepackage{graphicx} 
\usepackage{amsmath} 
\usepackage{wasysym} 

\newcommand{\vv}[1]{\mathbf{#1}} 
\newcommand{\vg}[1]{\boldsymbol{#1}}

\newcommand{\bra}[1]{\left< #1 \right|} 
\newcommand{\ket}[1]{\left| #1 \right>} 
\newcommand{\braket}[2]{\left< #1 \middle| #2 \right>} 
\newcommand{\brakett}[2]{\langle #1 | #2 \rangle} 

\newcommand{\divv}[1]{\mathop{\mathrm{div}} #1} 
\newcommand{\rot}[1]{\mathop{\mathrm{rot}} #1} 

\ifx \myStyle \styleJOSAB
\renewenvironment{multline}{\begin{equation}}{\end{equation}}
\fi

\begin{document}

\title{Coupled mode theory for on-channel nonlinear microcavities}

\ifx \myStyle \stylePRA
\author{Victor Grigoriev}
\email[]{victor.grigoriev@mpl.mpg.de}
\author{Fabio Biancalana}
\affiliation{Max Planck Institute for the Science of Light, G\"{u}nther-Scharowsky-Strasse 1, Bau 26, Erlangen D-91058, Germany}
\date{\today}
\fi

\ifx \myStyle \styleJOSAB
\author{Victor Grigoriev$^*$ and Fabio Biancalana}
\address{Max Planck Institute for the Science of Light, G\"{u}nther-Scharowsky-Strasse 1, Bau 26, Erlangen D-91058, Germany}
\address{$^*$Corresponding author: victor.grigoriev@mpl.mpg.de}
\fi

\begin{abstract}
We consider a nonlinear microcavity separating a waveguide channel into two parts so as the coupling between them is possible only due to the resonant properties of the microcavity. We provide a rigorous derivation of the equations used in the phenomenological coupled mode theory for such systems. This allows us to find the explicit formulas for all fitting parameters such as decay rates, coupling coefficients and characteristic intensities in terms of the mode profiles. The advantages of using the semi-analytical approach are discussed, and the accuracy of the results is compared with the strictly numerical methods. A particular attention is paid to multilayered structures since they represent the simplest realization of on-channel microcavities.
\end{abstract}

\pacs{290.5825, 140.3945, 190.3270, 190.1450, 230.5298, 230.4170.}

\maketitle

\section{Introduction\label{sIntroduction}}

The phenomenological Coupled Mode Theory (CMT) has been widely applied in optoelectronics, photonics and quantum optics to describe the linear and nonlinear properties of resonators \cite{Haus1984, Joannopoulos2008, Walls2008}. It played an importation role in the development of devices which can be embedded into photonic crystals: from waveguide splitters and add-drop filters \cite{Fan2001, Fan1998} to optical diodes and transistors \cite{Grigoriev2011a, Yanik2003}. Moreover, it was used to achieve an efficient generation of harmonics and difference frequencies in microcavities \cite{Hashemi2009, Burgess2009}, to describe bistable and multistable switching \cite{Soljacic2002, Grigoriev2010}, and to explain self-pulsations and chaotic behavior in coupled microcavities \cite{Grigoriev2011, Maes2009}. The primary advantage of CMT comes from the fact that it allows one to understand properly the physical interactions between different modes of a system which are often hidden in the strictly numerical simulations.

The CMT equations can be obtained from general physical concepts like
conservation of energy and time-reversal symmetry \cite{Fan2003, Suh2004}. As a consequence, these equations contain several fitting parameters: decay rates of resonances, coupling coefficients to different scattering channels and characteristic powers which describe the strength of the nonlinear effects. These parameters can be extracted from the experimental data, but most often they are determined after performing additional simulations in the time domain \cite{Soljacic2004, Bravo-Abad2007}. The main goal of this paper is to show that the CMT equations for the three-dimensional microcavities can be derived directly from the Maxwell equations without resorting to the phenomenological concepts. As a result, we were able to obtain the explicit formulas for all fitting parameters in terms of the mode profiles. We restrict our attention to the on-channel (or resonantly coupled) microcavities as opposed to the off-channel (side-coupled) microcavities \cite{Maes2005}, since the former case can be readily applied to multilayered structures, and the results can be simplified considerably.

The paper is organized as follows. In Section~\ref{sMaxwell}, we formulate the eigenvalue problem for a microcavity with two coupling ports and construct a complete set of orthogonal modes to expand an arbitrary field in it. To treat the electric and magnetic fields on equal footing, the Maxwell equations are written in a form which is similar to the Schr\"{o}dinger equation. It is emphasized that due to the time-reversal symmetry the Maxwell equations are doubly degenerate, and two fundamental modes exist for any resonant frequency. In Section~\ref{sEnvelopes}, we use these modes as a basis to describe the behavior of the microcavity in the vicinity of resonance. The CMT equations are derived for both cases when the fundamental modes are represented by traveling and standing waves. It is shown that the standing waves basis is particularly suitable for microcavities of high quality factors. Section \ref{sKerr} explains how to take into account perturbations caused by the Kerr nonlinearity and how to define the transfer matrix for the nonlinear microcavities. It provides the full set of equations for the time domain and frequency domain simulations including the explicit formulas for all fitting parameters. Section \ref{sExamples} presents several numerical examples and compares the accuracy of the CMT equations with other methods. It is also demonstrated how to apply the CMT equations to describe the nonlinear properties of microcavities with several localization centers.

\section{Generalized eigenvalue problem for Maxwell's equations\label{sMaxwell}}
\subsection{Analogy with the Schr\"{o}dinger equation\label{sMaxwell`sSchroedinger}}
\begin{figure}[b]
\center{\includegraphics[width=80mm]{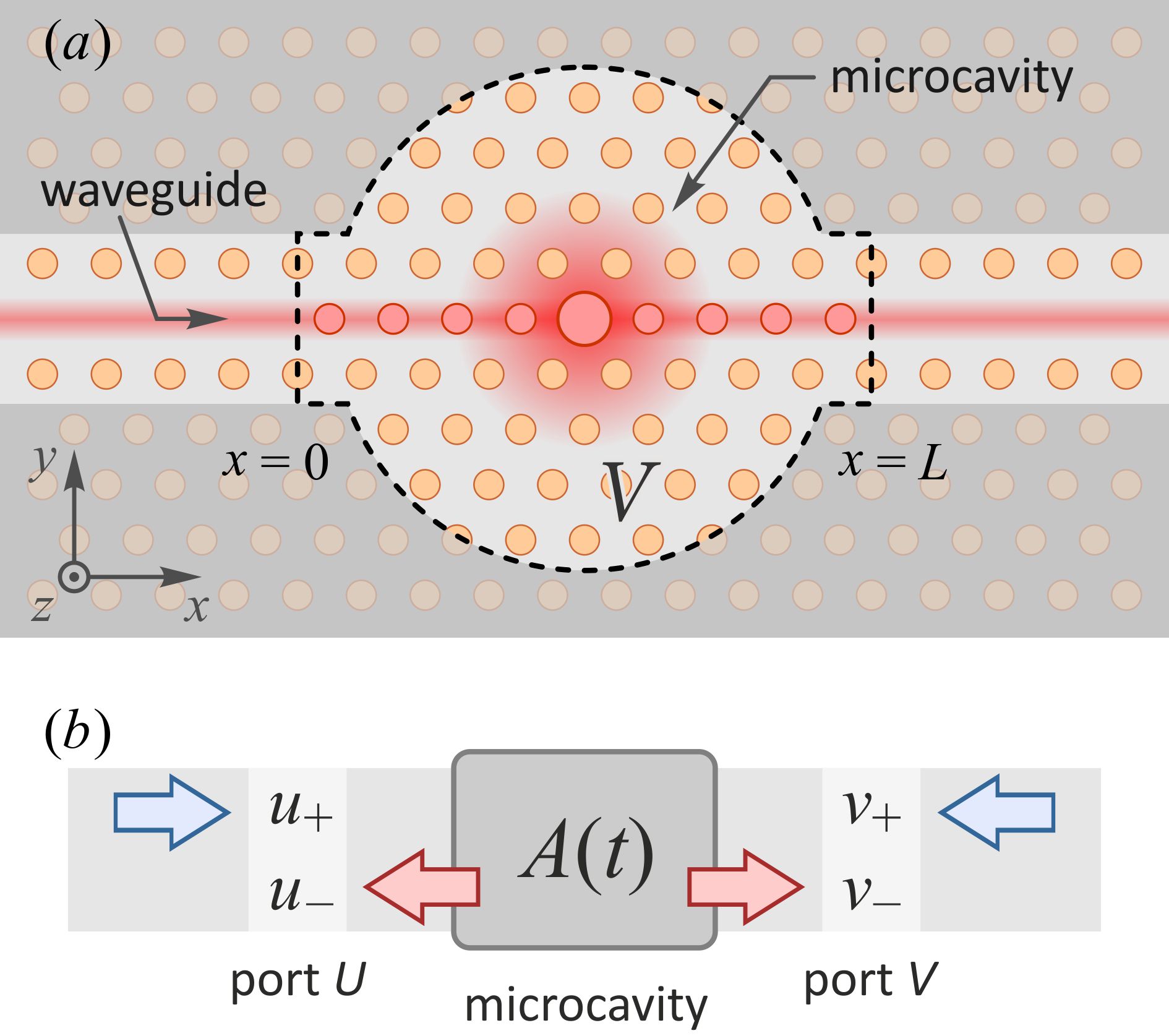}}
\caption{\label{figGeometry}
(Color online).
(a)~Photonic crystal waveguide with an embedded defect which behaves as a microcavity. The region where the electromagnetic filed is negligible has a dark gray background. The integration volume for the eigenvalue problem is shown by the dashed line. The dominant flow of energy goes through the reference planes at $x = 0$ and $x = L$.
(b)~A sketch of the model used in the scattering formalism. The amplitudes of the ingoing waves ($u_+$ and $v_+$) are related to the outgoing waves ($u_-$ and $v_-$) through the amplitude of the microcavity mode ($A$).
}
\end{figure}

To work with the electric and magnetic fields on equal footing, the Maxwell equations
\begin{gather}
\label{eRotE}
c \nabla \times \vv{E} = - \mu  \partial_t \vv{H}, \\
\label{eRotH}
c \nabla \times \vv{H} = \varepsilon \partial_t \vv{E},
\end{gather}
where $\varepsilon$ is permittivity, $\mu$ is permeability, and $c$ is the speed of light in the vacuum, can be rewritten in a form which is similar to the Schr\"{o}dinger equation
\begin{equation}
\label{eSchroedinger}
i \hbar \hat{\vg{\rho}} \partial_t \ket{\Psi} = \hat{\vv{L}} \ket{\Psi}.
\end{equation}
The wave function $\ket{\Psi}$ combines the electric and magnetic fields in a single vector
\begin{equation}
\label{eVector}
\ket{\Psi} = \left( \begin{array}{c}
   \vv{E}  \\
   \vv{H}  \\
\end{array} \right),
\end{equation}
with the inner product defined as
\begin{equation}
\label{eNorm}
\braket{\Psi_a}{\Psi_b} =
\iiint(
\vv{E}_a^* \cdot \vv{E}_b^{\phantom*} +
\vv{H}_a^* \cdot \vv{H}_b^{\phantom*}
)\mathrm{d}V,
\end{equation}
where the integration is performed over the volume of the microcavity as shown in Fig.~\ref{figGeometry}(a).

The operator $\hat{\vg{\rho}}$ is used as a weighting function, and it takes the following form for isotropic materials
\begin{equation}
\label{eRo}
\hat{\vg{\rho}} = \left[ \begin{array}{cc}
   \varepsilon\hat{I} & 0  \\
   0 & \mu\hat{I}  \\
\end{array} \right],
\end{equation}
where $\hat{I}$ is the identity tensor. The generalization to the case of more complicated constitutive relations such as those present in anisotropic or bi-anisotropic media is straightforward \cite{Serdyukov2001, Tuz2011}. It is important that the weighting operator is Hermitian ($\hat{\vg{\rho}}^\dagger = \hat{\vg{\rho}}$) for lossless media.

The operator $\hat{\vv{L}}$ can be interpreted as a Hamiltonian and is defined as
\begin{equation}
\label{eL}
\hat{\vv{L}} = i \hbar c \left[ {\begin{array}{cc}
   0 & \nabla \times  \\
   -\nabla \times & 0  \\
\end{array}} \right].
\end{equation}
To prove that this operator is Hermitian, it is sufficient to show that $
\brakett{\Psi_a}{\hat{\vv{L}} \Psi_b} = \brakett{\hat{\vv{L}} \Psi_a}{\Psi_b}
$ for two arbitrary solutions with the same boundary conditions. This relation can be transformed to a surface integral around the microcavity [Fig.~\ref{figGeometry}(a)] by using the identity $
\divv(\vv{A} \times \vv{B}) = \vv{B} \rot\vv{A} - \vv{A} \rot\vv{B}
$
\begin{equation}
\label{eHermitian}
\oiint(
{\vv{E}}_a^* \times {\vv{H}}_b^{\phantom*} +
{\vv{E}}_b^{\phantom*} \times {\vv{H}}_a^*
)\mathrm{d}{\vv{s}} = 0.
\end{equation}
The integral is nonzero only at the input and output ports so that it can be reduced to
\begin{equation}
\label{eBoundary}
(
{\vv{E}}_a^* \times {\vv{H}}_b^{\phantom*} +
{\vv{E}}_b^{\phantom*} \times {\vv{H}}_a^*
) |_{x = 0}^{x = L} = 0.
\end{equation}
This equation is always valid for solutions that satisfy the periodic (Bloch) boundary conditions at the coupling ports
\begin{equation}
\label{eBloch}
\ket{\Psi(x = L)} = {\textrm{e}}^{i \varphi} \ket{\Psi(x = 0)},
\end{equation}
where $\varphi$ is a real constant.

\subsection{Orthogonality relations between modes\label{sMaxwell`sEigenproblem}}
Assuming the time dependence of the form $
\ket{\Psi(\vv{r},t)} = \ket{\psi(\vv{r})} \textrm{e}^{-i \omega t},
$ the generalized eigenvalue problem for Eq.~(\ref{eSchroedinger}) can be formulated as
\begin{equation}
\label{eEigenproblem}
\hat{\vv{L}} \ket{\psi_m} =
\hbar \omega_m
\hat{\vg{\rho}} \ket{\psi_m}.
\end{equation}
Since both operators $\hat{\vv{L}}$ and $\hat{\vv{\rho}}$ are Hermitian, the eigenvalues, or resonant frequencies $\omega_m$, should be real. Moreover, the eigenvectors, or modes corresponding to different frequencies $\omega_n \ne \omega_m$, should satisfy the orthogonality relation
\begin{equation}
\label{eOrthogonality}
\braket {\psi_n} {\hat{\vg{\rho}} \psi_m} =
\iiint{(
\varepsilon {\vv{E}}_n^* \cdot {\vv{E}}_m +
\mu {\vv{H}}_n^* \cdot {\vv{H}}_m
)\mathrm{d}V} = 0.
\end{equation}
These modes can be normalized to a volume of the microcavity, but we prefer to preserve their meaning as the total energy stored in the mode and will normalize the boundary conditions instead.

It is useful to introduce another Hermitian operator
\begin{equation}
\label{eSigma}
{\hat{\vg{\sigma}}}_{\!z}  =
\left[ \begin{array}{rr}
   \hat{I} & 0  \\
   0 &  -\hat{I}  \\
\end{array} \right],
\end{equation}
which satisfies the following commutation relations $
{\hat{\vg{\sigma}}}_{\!z} \hat{\vv{L}} =
-\hat{\vv{L}} {\hat{\vg{\sigma}}}_{\!z}
$
and $
{\hat{\vg{\sigma}}}_{\!z} \hat{\vg{\rho}} =
\hat{\vg{\rho}} {\hat{\vg{\sigma}}}_{\!z}
$. By using these properties, it can be shown that the solutions of Eq.~(\ref{eEigenproblem}) come in pairs, and for any solution with a positive frequency one can immediately construct another solution with a negative frequency
\begin{equation}
\label{eSigmaEigenproblem}
\hat{\vv{L}} \ket{{\hat{\vg{\sigma}}}_{\!z} \psi_m} = -
\hbar \omega_m
\hat{\vg{\rho}} \ket{{\hat{\vg{\sigma}}}_{\!z} \psi_m}.
\end{equation}
It is possible to return back to the positive frequencies by applying the operation of complex conjugation and to obtain two independent solutions for the same frequency~$\omega_m$
\begin{equation}
\label{eDegeneracy}
\ket{\psi_m^+} = \left( {\begin{array}{l}
   {\vv{E}}_m   \\
   {\vv{H}}_m   \\
\end{array}} \right)
\; \textrm{and} \;
\ket{\psi_m^-} = \left( {\begin{array}{l}
   \phantom{-} {\vv{E}}_m^*  \\
   -{\vv{H}}_m^*  \\
\end{array}} \right).
\end{equation}
In homogeneous media, these solutions can be considered as plane waves moving in the opposite directions $(E_y, \; \pm H_z)^\mathrm{T} \exp [i (\pm k_x x - \omega t)]$.

Similar to the derivation of the orthogonality relations, it can be proved that for $\omega_n  \ne - \omega_m$
\begin{equation}
\label{eSigmaOrthogonality}
\braket {\psi_n} {{\hat{\vg{\sigma}}}_{\!z}  \hat{\vg{\rho}} \psi_m} =
\iiint(
\varepsilon {\vv{E}}_n^* \cdot {\vv{E}}_m^{\phantom*} -
\mu {\vv{H}}_n^* \cdot {\vv{H}}_m^{\phantom*}
)\mathrm{d}V = 0.
\end{equation}
For modes of the same frequency $\omega_n  = \omega_m$, it shows that the averaged electric energy stored in the mode is equal to the magnetic energy $
\iiint {\varepsilon |{\vv{E}}_m |^2 \mathrm{d}V} =
\iiint {\mu |{\vv{H}}_m|^2 \mathrm{d}V}
$. If Eqs.~(\ref{eOrthogonality}) and (\ref{eSigmaOrthogonality}) are combined, the orthogonality relations can be separately formulated for the electric and magnetic fields ($\omega_n \ne \pm \omega _m$)
\begin{equation}
\label{eCombinedOrthogonality}
\iiint {\varepsilon {\vv{E}}_n^* \cdot {\vv{E}}_m^{\phantom*} \mathrm{d}V} =
\iiint {\mu {\vv{H}}_n^* \cdot {\vv{H}}_m^{\phantom*} \mathrm{d}V} = 0.
\end{equation}

\section{Slowly varying envelopes\label{sEnvelopes}}
\subsection{Traveling waves basis\label{sEnvelopes`sTraveling}}
In the vicinity of the resonant frequency $\omega_m$, the field can be searched in the following form
\begin{equation}
\label{eProbeT}
\ket{\Psi} =
a_+(\vv{r},t) \ket{\psi_m^+} {\mathrm{e}}^{ - i \omega_m t} +
a_-(\vv{r},t) \ket{\psi_m^-} {\mathrm{e}}^{ - i \omega_m t},
\end{equation}
where $a_\pm(\vv{r},t)$ are slowly varying envelopes of forward and backward moving waves. Substituting the approximate solution (\ref{eProbeT}) into the Schr\"{o}dinger equation (\ref{eSchroedinger}) gives
\begin{multline}
\label{eSchroedingerT}
i \hbar \hat{\vg{\rho}} \left(
\partial_t a_+ \ket{\psi_m^+} +
\partial_t a_- \ket{\psi_m^-}
\right) = \\
\hat{\vv{L}} (a_+) \ket{\psi_m^+} +
\hat{\vv{L}} (a_-) \ket{\psi_m^-},
\end{multline}
where the notation $\hat{\vv{L}}(a_\pm)$ means that the derivatives in the operator $\hat{\vv{L}}$ are applied only to the scalar envelopes~$a_\pm$
\begin{equation}
\label{eLa}
\hat{\vv{L}}(a) = i \hbar c \left[ \begin{array}{cc}
   0 & (\nabla a) \times  \\
   - (\nabla a) \times  & 0  \\
\end{array} \right].
\end{equation}

Projecting Eq.~(\ref{eSchroedingerT}) on the vectors $\bra{\psi_m^\pm}$ gives a set of two coupled equations which describe the propagation of the envelopes
\begin{align}
\label{eEnvelopesT1}
\partial_t a_+ + ({\vv{v}}_{\mathrm{g}} \cdot \nabla ) a_+ &= -
g^* \partial_t a_-, \\
\label{eEnvelopesT2}
\partial_t a_- - ({\vv{v}}_{\mathrm{g}} \cdot \nabla ) a_- &= -
g \partial_t a_+.
\end{align}
The coupling terms appear due to the fact that the two fundamental modes $\ket{\psi_m^\pm}$ are not orthogonal
\begin{equation}
\label{eOverlapMP}
\braket{\psi_m^-} {\hat{\vg{\rho}} \psi_m^+} =
2 \iiint \varepsilon {\vv{E}}_m ^2 \mathrm{d}V.
\end{equation}
It is convenient to measure their overlap by using their common norm
\begin{equation}
\label{eOverlapPP}
\braket{\psi_m^\pm} {\hat{\vg{\rho}} \psi_m^\pm} =
2 \iiint \varepsilon | {\vv{E}}_m |^2 \mathrm{d}V
\end{equation}
and to introduce a special parameter
\begin{equation}
\label{eGammaT}
g = \frac
{\iiint \varepsilon {\vv{E}}_m^2 \mathrm{d}V}
{\iiint \varepsilon |{\vv{E}}_m|^2 \mathrm{d}V}.
\end{equation}
Since the inequality $
\left| \int f(x) dx \right|^2 \le
\int \left| f(x) \right|^2 dx
$ holds for any complex function $f(x)$, the parameter $g$ is limited to the range $|g| \le 1$. It tends to zero in homogeneous media due to the rapid oscillation of phase in the numerator of Eq.~(\ref{eGammaT}). As a result, the coupling terms in Eqs.~(\ref{eEnvelopesT1})--(\ref{eEnvelopesT2}) disappear, and the two envelopes propagate independently. On the contrary, the field in nonhomogeneous media is localized around the defect regions, the amplitude varies very quickly in comparison to the phase, and this leads to large values of the parameter $g$. As a consequence, $g$ can describe the localization strength of a resonance.

In the derivation of Eqs.~(\ref{eEnvelopesT1})--(\ref{eEnvelopesT2}), it was also used that
\begin{gather}
\label{eOverlapPLP}
\!\!\! \braket{\psi_m^ \pm} {\hat{\vv{L}}(a) \psi_m^\pm} =
\mp 2i \hbar c \nabla a \cdot \iiint \mathop{\mathrm{Re}}({\vv{E}}_m^* \times {\vv{H}}_m^{\phantom{*}})\mathrm{d}V, \\
\label{eOverlapPLM}
\braket{\psi_m^ \pm} {\hat{\vv{L}}(a) \psi_m^\mp} = 0,
\end{gather}
and the parameter ${\vv{v}}_{\mathrm{g}}$ was introduced
\begin{equation}
\label{eVg}
{\vv{v}}_{\mathrm{g}} =
c \frac
{\iiint \mathop{\mathrm{Re}} ({\vv{E}}_m^* \times {\vv{H}}_m^{\phantom{*}})\mathrm{d}V}
{\iiint \varepsilon |{\vv{E}}_m|^2 \mathrm{d}V}.
\end{equation}
The integrand in the numerator of Eq.~(\ref{eVg}) contains the averaged energy flow $
\vv{S}_W = (c / 8 \pi) \mathop{\mathrm{Re}} [{\vv{E}} \times {\vv{H}}^*]
$, which is a solenoidal vector field $\divv{\vv{S}_W} = 0$ for any resonant frequency, because the averaged energy density $
W = (1 / 16 \pi)(\varepsilon |{\vv{E}}|^2  + \mu |{\vv{H}}|^2 )
$ does not change in a stationary state. In multilayered structures, the energy flow depends only on the $x$-coordinate, which means that $
\divv{\vv{S}_W} = \partial_x (\vv{S}_W)_x  = 0
$, and as a result $
(\vv{S}_W)_x = \mathrm{const}
$ can be easily integrated. Therefore, the parameter ${\vv{v}}_{\mathrm{g}}$ can be interpreted as an effective group velocity. The same considerations should hold even for more complicated structures if the center of the microcavity is on the $x$-axis. Due to the symmetry, the dominant energy flow is also directed along the $x$-axis, and the directional derivative in Eqs.~(23)--(24) can be approximated as $
({\vv{v}}_{\mathrm{g}} \cdot \nabla) = (v_{\mathrm{g}} \partial_x)
$.

It is convenient to choose the position of reference planes in such a way that the boundary conditions (\ref{eBloch}) have $\varphi = 0$. Using these reference planes as channels for the in- and outgoing waves, it is possible to develop a scattering formalism [Fig.~\ref{figGeometry}(b)] and to find the transmission spectrum of the microcavity in the vicinity of the resonance $\omega _m$. Assuming that the reference planes are located at $x = 0$  and $x = L$, the boundary conditions for the forward and backward moving envelopes can be written as $a_\pm(0) = u_\pm$, $a_\pm(L) = v_\mp$. Approximating the spatial and temporal derivatives in Eqs.~(\ref{eEnvelopesT1})--(\ref{eEnvelopesT2}) with finite differences $\partial_x a_\pm = (v_ \mp - u_\pm) / L$ and $\partial_t a_\pm = - i \delta \omega (u_\pm + v_\mp) / 2$, one can show that the scattering matrix $\mathcal{S}_{uv}$ defined as
\begin{equation}
\label{eSuvDefinition}
\begin{pmatrix}
   u_-  \\
   v_-  \\
\end{pmatrix} =
\mathcal{S}_{uv}
\begin{pmatrix}
   u_+  \\
   v_+  \\
\end{pmatrix} =
\left[ \begin{array}{cc}
   r_u & t  \\
   t & r_v  \\
\end{array} \right]
\begin{pmatrix}
   u_+  \\
   v_+  \\
\end{pmatrix},
\end{equation}
up to the first order of the detuning from the resonance $\delta \omega = \omega  - \omega_m$ is
\begin{equation}
\label{eSuvResult}
\mathcal{S}_{uv} = \frac{1}{1 - i (\delta \omega / \gamma)}
\left[ {\begin{array}{cc}
   ig (\delta \omega / \gamma) & 1  \\
   1 & ig^* (\delta \omega / \gamma)  \\
\end{array}} \right],
\end{equation}
where $\gamma = v_{\mathrm{g}} / L$ or
\begin{equation}
\label{eGamma3D}
\gamma = \frac {c \sigma} {\iiint \varepsilon |{\vv{E}}_m|^2 \mathrm{d}V},
\end{equation}
which is obtained from Eq.~(\ref{eVg}) in the assumption that the mode profiles are considered as dimensionless quantities and normalized in such a way that
\begin{equation}
\label{eSigma}
\sigma = \iint \mathop{\mathrm{Re}} [({\vv{E}}_m^* \times {\vv{H}}_m^{\phantom{*}})_x] \mathrm{d}y \mathrm{d}z = 1 \; \mathrm{cm}^2.
\end{equation}

The transmission $t(\omega )$ and reflection $r_{u,v}(\omega)$ spectra for waves incident on the ports $U$ or $V$ can be found by direct comparison of the matrix elements in Eqs.~(\ref{eSuvDefinition}) and (\ref{eSuvResult}). For example, the transmission coefficient is given by
\begin{equation}
\label{eLinearTransmission}
t(\omega) =
\left[ {1 - i (\delta \omega / \gamma)} \right]^{-1},
\end{equation}
and thus the resonance contour has the Lorentzian shape with the half-width at half-maximum equal to $\gamma$.

\subsection{Standing waves basis\label{sEnvelopes`sStanding}}
\begin{figure}[b]
\center{\includegraphics[width=80mm]{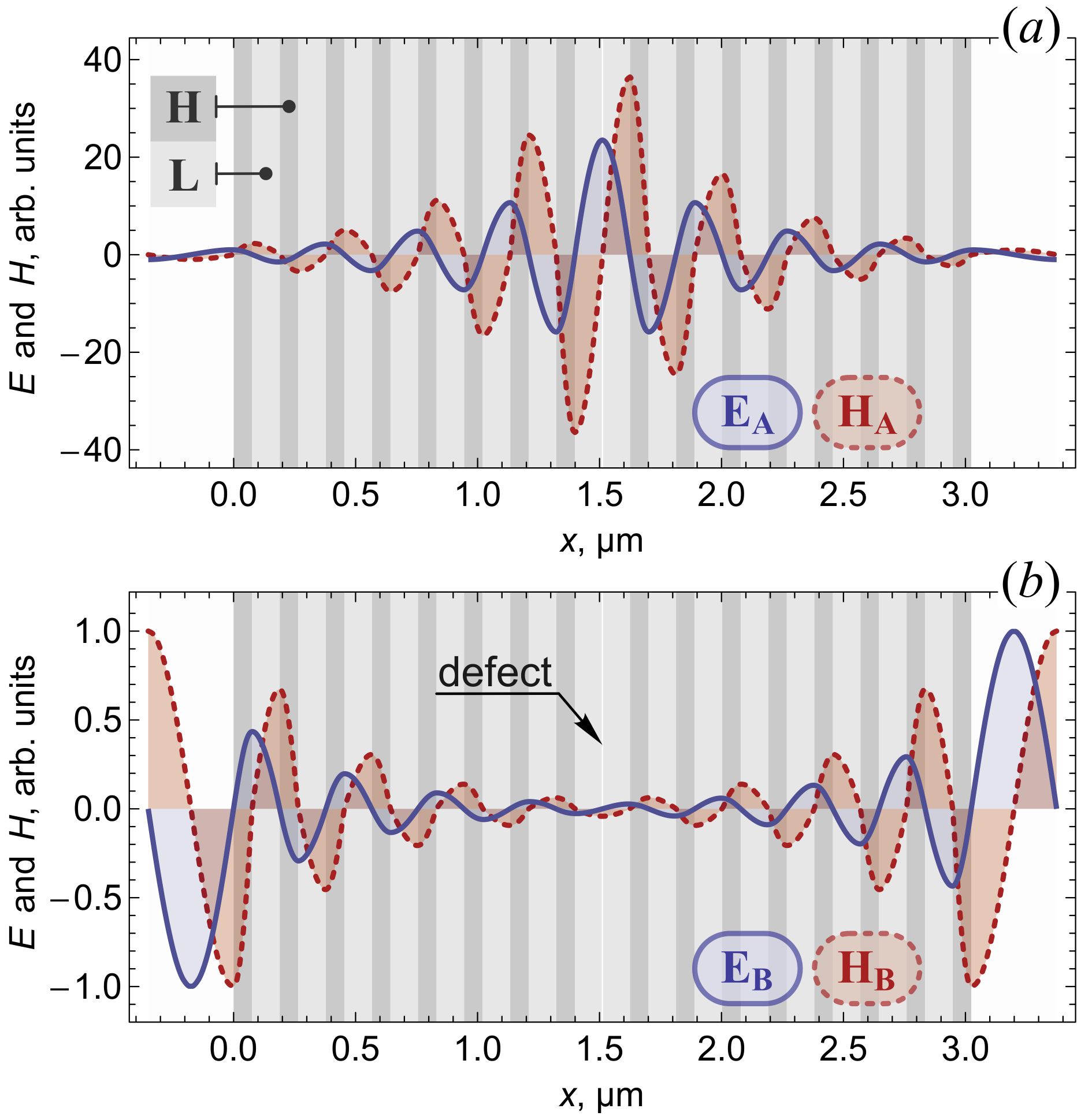}}
\caption{\label{figModes}(Color online).
An example of two fundamental solutions that exist at the same resonance frequency and satisfy the boundary conditions of standing waves with zero energy flow. The electric and magnetic fields are shown by solid and dashed lines, respectively. For one of the solutions they are exponentially growing towards the defect region in the middle~(a) while for the other independent solution they are exponentially decaying~(b). Background shows alternation of layers with a higher ('H') and lower ('L') index of refraction inside the structure.
}
\end{figure}

As was mentioned before, the basis formed by two modes $\ket{\psi_m^\pm}$ is not orthogonal. However, a linear combination of these modes can be used to construct a new basis which will have such a property. This is particularly easy to do for mirror symmetric structures
\begin{gather}
\label{ePsiA}
\ket{\psi_{\mathrm{A}}} = \!\left( \begin{array}{l}
   {\vv{E}}_{\mathrm{A}}  \\
   {\vv{H}}_{\mathrm{A}}  \\
\end{array} \!\right)\!  =
\frac{\ket{\psi_m^+} + \ket{\psi_m^-}}{2} =
\!\left(\! \begin{array}{l}
   \phantom{i} \mathop{\mathrm{Re}} [{\vv{E}}_m]  \\
   i \mathop{\mathrm{Im}} [{\vv{H}}_m]  \\
\end{array} \right)\!, \\
\label{ePsiB}
\ket{\psi_{\mathrm{B}}} = \!\left( \begin{array}{l}
   {\vv{E}}_{\mathrm{B}}  \\
   {\vv{H}}_{\mathrm{B}}  \\
\end{array} \!\right)\!  =
\frac{\ket{\psi_m^+} - \ket{\psi_m^-}} {2i} =
\!\left(\! \begin{array}{l}
   \phantom{-i} \mathop{\mathrm{Im}} [{\vv{E}}_m]  \\
    -i \mathop{\mathrm{Re}} [{\vv{H}}_m ]  \\
\end{array} \right)\!.
\end{gather}
The new modes $\ket{\psi_{\mathrm{A}}}$ and $\ket{\psi_{\mathrm{B}}}$ correspond to standing waves because the energy flow ${\vv{S}}_W$  for them is zero by definition. It can be shown that when one of them is exponentially growing around a defect region, the other mode is exponentially decaying [Fig.~\ref{figModes}]. As a result, the norms of these modes can differ significantly for resonances with strong localization. By using the property $
\ket{\psi_m^\pm} =
\ket{\psi_{\mathrm{A}}} \pm i
\ket{\psi_{\mathrm{B}}}
$, the parameter $g$ in Eq.~(\ref{eGammaT}) can be rewritten as
\begin{equation}
\label{eGammaS}
g = \frac
{\iiint \varepsilon ({\vv{E}}_{\mathrm{A}}^2 - {\vv{E}}_{\mathrm{B}}^2)\mathrm{d}V}
{\iiint \varepsilon ({\vv{E}}_{\mathrm{A}}^2 + {\vv{E}}_{\mathrm{B}}^2)\mathrm{d}V}.
\end{equation}
Therefore, $g$ is real for mirror symmetric structures and tends to $\pm 1$ depending on which mode dominates.

The total field can be expanded in the new basis as
\begin{equation}
\label{eProbeS}
\ket{\Psi} =
A(\vv{r},t)\ket{\psi_{\mathrm{A}}} {\mathrm{e}}^{ - i \omega_m t} +
i B (\vv{r},t)\ket{\psi_{\mathrm{B}}} {\mathrm{e}}^{ - i \omega_m t},
\end{equation}
where $A(\vv{r},t)$ and $B(\vv{r},t)$ are slowly varying amplitudes of the modes. The equations which describe their propagation can be found by substituting Eq.~(\ref{eProbeS}) into Eq.~(\ref{eSchroedinger})
\begin{multline}
\label{eSchroedingerS}
i \hbar \hat{\vg{\rho}} \left(
\partial_t A \ket{\psi_{\mathrm{A}}} +
i \partial_t B \ket{\psi_{\mathrm{B}}}
\right) = \\
\hat{\vv{L}}(A) \ket{\psi_{\mathrm{A}}} +
i \hat{\vv{L}}(B) \ket{\psi_{\mathrm{B}}}.
\end{multline}
Making projection of this equation on $\bra{\psi_{\mathrm{A}}}$ and then on $\bra{\psi_{\mathrm{B}}}$ leads to
\begin{align}
\label{eEnvelopesS1}
(1 + g) \partial_t A +
({\vv{v}}_{\mathrm{g}} \cdot \nabla ) B &= 0,  \\
\label{eEnvelopesS2}
(1 - g) \partial_t B +
({\vv{v}}_{\mathrm{g}} \cdot \nabla ) A &= 0,
\end{align}
where the following properties were used
\begin{gather}
\label{eOverlapAA}
\braket{\psi_{\mathrm{A}}} {\hat{\vv{L}}(A)\psi_{\mathrm{A}}} =
\braket{\psi_{\mathrm{B}}} {\hat{\vv{L}}(B)\psi_{\mathrm{B}}} = 0, \\
\label{eOverlapBA}
\braket{\psi_{\mathrm{B}}} {\hat{\vv{L}}(A)\psi_{\mathrm{A}}} =
\hbar c \nabla A \cdot \iiint \mathop{\mathrm{Re}}
({\vv{E}}_m^* \times {\vv{H}}_m^{\phantom *}) \mathrm{d}V, \\
\label{eOverlapAB}
\braket{\psi_{\mathrm{A}}} {\hat{\vv{L}}(B)\psi_{\mathrm{B}}} =
- \hbar c \nabla B \cdot \iiint \mathop{\mathrm{Re}}
({\vv{E}}_m^* \times {\vv{H}}_m^{\phantom *}) \mathrm{d}V.
\end{gather}

The Eqs.~(\ref{eEnvelopesS1})--(\ref{eEnvelopesS2}) can be applied to develop the scattering formalism in the standing waves basis. The boundary conditions can be obtained by using the following relations with the traveling waves basis: $A = a_+ + a_-$ and $B = a_+ - a_-$. It is convenient to consider $A$ and $B$ as a function of time only by taking their average value on the boundaries
\begin{gather}
\label{eAaveraged}
A = \frac {A(0) + A(L)} {2} =
\frac {(u_+ + u_-) + (v_- + v_+)} {2}, \\
\label{eBaveraged}
B = \frac {B(0) + B(L)} {2} =
\frac {(u_+ - u_-) + (v_- - v_+)} {2},
\end{gather}
which can be written in the matrix from as
\begin{equation}
\label{eScatteringAB}
\begin{pmatrix}
   u_-  \\
   v_-  \\
\end{pmatrix} = -
\left[ \begin{array}{cc}
   0 & 1  \\
   1 & 0  \\
\end{array} \right]
\begin{pmatrix}
   u_+  \\
   v_+  \\
\end{pmatrix} + A
\left( \begin{array}{c}
   1  \\
   1  \\
\end{array} \right) + B
\begin{pmatrix}
   -1  \\
   1   \\
\end{pmatrix}.
\end{equation}
The spatial derivatives can be approximated then by
\begin{align}
\label{eAderivative}
\frac {\partial A} {\partial x} &=
\frac {A(L) - A(0)} {L} =
2 \frac {B - (u_+ - v_+)} {L}, \\
\label{eBderivative}
\frac {\partial B} {\partial x} &=
\frac {B(L) - B(0)} {L} =
2 \frac {A - (u_+ + v_+)} {L}.
\end{align}
Therefore, Eqs.~(\ref{eEnvelopesS1})--(\ref{eEnvelopesS2}) take the following form
\begin{align}
\label{eAmplitudeA}
\frac{1 + g}{2} \frac{\mathrm{d}A}{\mathrm{d}t} &= - \gamma A + \gamma (u_+ + v_+), \\
\label{eAmplitudeB}
\frac{1 - g}{2} \frac{\mathrm{d}B}{\mathrm{d}t} &= - \gamma B + \gamma (u_+ - v_+).
\end{align}
where the decay rate $\gamma$ is given by
\begin{equation}
\label{eGamma}
\gamma  = \frac
{c \sigma}
{\iiint \varepsilon ({\vv{E}}_{\mathrm A}^2 + {\vv{E}}_{\mathrm B}^2) \mathrm{d}V}.
\end{equation}

A considerable simplification can be made for resonances with large quality factors. In this case, the parameter $g$ tends to $\pm 1$  depending on which mode dominates. If the even mode has a larger norm $
\iiint {\varepsilon {\vv{E}}_{\mathrm{A}}^2 \mathrm{d}V} \gg
\iiint {\varepsilon {\vv{E}}_{\mathrm{B}}^2 \mathrm{d}V}
$, then according to Eq.~(\ref{eGammaS}) $g \to 1$ and Eq.~(\ref{eAmplitudeB}) reduces to $B = u_+ - v_+$. The full set of the CMT equations is thus
\begin{gather}
\label{eCMT`Even`Lin1}
\frac{\mathrm{d}A}{\mathrm{d}t} =  - \gamma A + \gamma (u_+ + v_+), \\
\label{eCMT`Even`Lin2}
\begin{pmatrix}
 u_-   \\
 v_-   \\
\end{pmatrix} = -
\begin{pmatrix}
 u_+   \\
 v_+   \\
\end{pmatrix} + A
\left( \begin{array}{c}
   1  \\
   1  \\
\end{array} \right).
\end{gather}
On the contrary, if the odd mode has a larger norm, then $g \to  -1$ and Eq.~(\ref{eAmplitudeA}) reduces to $A = u_+ + v_+$. The full set of the CMT equations is thus
\begin{gather}
\label{eCMT`Odd`Lin1}
\frac{\mathrm{d}B}{\mathrm{d}t} =  - \gamma B + \gamma (u_+ - v_+), \\
\label{eCMT`Odd`Lin2}
\begin{pmatrix}
 u_-   \\
 v_-   \\
\end{pmatrix} =
\begin{pmatrix}
 u_+   \\
 v_+   \\
\end{pmatrix} + B
\begin{pmatrix}
   -1  \\
   1  \\
\end{pmatrix}.
\end{gather}

\section{Perturbations caused by the Kerr nonlinearity\label{sKerr}}
\subsection{Time domain\label{sKerr`sTime}}
The nonlinear effects can be treated as perturbation terms in Eq.~(\ref{eSchroedinger}), and the standing waves basis is particularly suitable for this purpose because the dynamics of the system can be described only by one variable. The left hand side of Eq.~(\ref{eSchroedingerS}) should be modified to include the perturbation term
\begin{equation}
\label{ePerturbationTerm}
\hbar \omega_m \Delta \hat{\vg{\rho}} \left(
A \ket{\psi_{\mathrm{A}}} + iB \ket{\psi_{\mathrm{B}}}
\right).
\end{equation}
As a source of $\Delta \hat{\vg{\rho}}$, the Kerr nonlinearity will be considered, which corresponds to the following constitutive relation between the electric field and the displacement vector
\begin{equation}
\label{eDisplacement}
{\vv{D}} = \varepsilon \vv{E} + \varepsilon_{\mathrm{K}} {\vv{E}}^3.
\end{equation}
If the effect of the third harmonic generation, which is represented by the first term in the expansion
\begin{multline}
\label{eHarmonics}
(\mathop{\mathrm{Re}} [{\vv{E}}_\omega {\mathrm{e}}^{- i \omega t}])^3 = \\
(1 / 4) \mathop{\mathrm{Re}}
[{\vv{E}}_\omega^3 {\mathrm{e}}^{-3i \omega t}] +
(3 / 4) |{\vv{E}}_\omega|^2 \mathop{\mathrm{Re}}
[{\vv{E}}_\omega {\mathrm{e}}^{- i \omega t}],
\end{multline}
can be neglected, the constitutive relation is reduced to
\begin{equation}
\label{eDisplacementSimplified}
\vv{D} = \varepsilon \vv{E} +
(3 / 4) \varepsilon_{\mathrm{K}} |\vv{E}|^2 \vv{E},
\end{equation}
which means that
\begin{equation}
\label{eDeltaRho}
\Delta \hat{\vg{\rho}} = \frac{3}{4}
\left[ \begin{array}{cc}
   \varepsilon_{\mathrm{K}} |\vv{E}|^2 \hat{I} & 0  \\
   0 & 0  \\
\end{array} \right].
\end{equation}
Since the electric field can be written as $\vv{E} = A{\vv{E}}_{\mathrm{A}} + i B {\vv{E}}_{\mathrm{B}}
$, the projection of Eq.~(\ref{ePerturbationTerm}) on $\bra{\psi_{\mathrm{A}}}$ and then on $\bra{\psi_{\mathrm{B}}}$ leads to the overlap integrals of the following form
\begin{equation}
\label{eGammaCoefficients}
\Gamma_{{\mathrm{A}}^p {\mathrm{B}}^q} =
\frac{3 \omega_m}{4}
\frac
{\iiint \varepsilon_{\mathrm{K}} ({\vv{E}}_{\mathrm{A}})^p ({\vv{E}}_{\mathrm{B}})^q \mathrm{d}V}
{\iiint \varepsilon ({\vv{E}}_{\mathrm{A}}^2 + {\vv{E}}_{\mathrm{B}}^2)\mathrm{d}V},
\end{equation}
where $p$ and $q$ are nonnegative integers with an additional restriction $p + q = 4$. There is only a small number of nonzero coefficients $\Gamma$ which should be taken into account due to the fact that ${\vv{E}}_{\mathrm{A}}$ and ${\vv{E}}_{\mathrm{B}}$ are functions of the opposite parity.
Eqs.~(\ref{eEnvelopesS1})--(\ref{eEnvelopesS2}) in presence of the Kerr nonlinearity are consequently
\begin{align}
\label{eEnvelopesNonlinear1}
(1 + g) &\partial_t A + ({\vv{v}}_{\mathrm{g}} \cdot \nabla) B = \ifx \myStyle \stylePRA \nonumber \\ \fi
i A |A|^2 \Gamma_{\mathrm{AAAA}} + i ( 2 A |B|^2 - A^* B^2 ) \Gamma_{\mathrm{AABB}}, \\
\label{eEnvelopesNonlinear2}
(1 - g) &\partial_t B + ({\vv{v}}_{\mathrm{g}} \cdot \nabla) A = \ifx \myStyle \stylePRA \nonumber \\ \fi
i B |B|^2 \Gamma_{\mathrm{BBBB}} + i ( 2 B |A|^2 - B^* A^2 ) \Gamma_{\mathrm{AABB}}.
\end{align}
The nonlinear effects described by these equations include the self-phase modulation (terms proportional to $A |A|^2$ and $B |B|^2$) as well as the cross-phase modulation ($A |B|^2$ and $B |A|^2$). If the coefficients $\Gamma$ were equal, the relative strength of these effects would be given by the factor of 2, which coincides with the results obtained in the case of shallow gratings \cite{Agrawal2008}. It is worth noting that there are a few additional terms ($A^* B^2$ and $B^* A^2$), which are responsible for the phase conjugation and do not exist in the shallow gratings \cite{Sterke1996}.

It turns out however that only the self-phase modulation is important for resonances with large quality factors. If the even mode $\ket{\psi_{\mathrm{A}}}$ dominates, the overlap integrals differ significantly $\Gamma_{\mathrm{AAAA}} \gg \Gamma_{\mathrm{AABB}} \gg \Gamma_{\mathrm{BBBB}}$. Therefore, to take into account the Kerr nonlinearity the CMT equations (\ref{eCMT`Even`Lin1}) and (\ref{eCMT`Odd`Lin1}) should be modified in the following way
\begin{gather}
\label{eCMT`Even`Nonlinear}
\frac{\mathrm{d}A}{\mathrm{d}t} =  - \left[ i \left(
\omega_m  - \gamma \frac{|A|^2}{I_{\mathrm{A}}}
\right) + \gamma \right] A +
\gamma (u_+ + v_+), \\
\label{eCMT`Odd`Nonlinear}
\frac{\mathrm{d}B}{\mathrm{d}t} =  - \left[ i \left(
\omega_m  - \gamma \frac{|B|^2}{I_{\mathrm{B}}}
\right) + \gamma \right] B +
\gamma (u_+ - v_+).
\end{gather}
The main influence of the Kerr nonlinearity results in the shift of the resonant frequency. To emphasize this, the time dependence $\mathrm{exp}( - i \omega_m t)$ used as a factor in Eq.~(\ref{eProbeS}) was explicitly included in the amplitudes $A$ and $B$. Two new parameters were introduced $I_{\mathrm{A}}$ and $I_{\mathrm{B}}$ which have the meaning of characteristic intensities. Before giving the explicit formulas for them, it is important to choose proper units for the electric field.

The intensities will be measured in $\textrm{MW} / \textrm{cm}^2$, and the nonlinear refractive index $n_2$ caused by the Kerr nonlinearity will be specified in $\textrm{cm}^2 / \textrm{MW}$ for consistency. The transition from the intensity dependent refractive index $n(I) = n + n_2 I$ to the permittivity can be performed as $
\varepsilon (E) = \varepsilon  + 2 \varepsilon n_2 |E|^2$, where the units for the electromagnetic field were defined so as to produce a unit energy flow in vacuum, namely $
I_{\mathrm{unit}} = (c / 8 \pi)
|E_{\mathrm{unit}}|^2  = 1 \; \mathrm{MW} / {\mathrm{cm}}^2
$. Comparing with Eq.~(\ref{eDisplacementSimplified}) gives $ \varepsilon _{\mathrm{K}} = (8 / 3) \varepsilon n_2$
Therefore, the characteristic intensities are
\begin{equation}
\label{eCharacteristicPowers}
I_{\mathrm{A,B}} = \frac
{c \sigma}
{\omega_m \iiint \varepsilon n_2
({\vv{E}}_{\mathrm{A,B}})^4 \mathrm{d}V}.
\end{equation}

\subsection{Frequency domain\label{sKerr`sFreq}}
It is worth noting that the CMT equations (\ref{eCMT`Even`Nonlinear}) and (\ref{eCMT`Even`Lin2}) for resonances with even modes (or Eqs.~(\ref{eCMT`Odd`Nonlinear}) and (\ref{eCMT`Odd`Lin2}) for odd modes) represent an extension of the scattering matrix method to the time domain. It is convenient to combine these equations
\begin{gather}
\label{eCMTmain1}
\frac{\mathrm{d}A}{\mathrm{d}t} =  - \left[ i \left(
\omega_0  - \gamma \frac {|A|^2} {I_0}
\right) + \gamma \right] A + \gamma (u_+ \pm v_+), \\
\label{eCMTmain2}
\begin{pmatrix}
 u_-  \\
 v_-  \\
\end{pmatrix} =  \mp
\begin{pmatrix}
 u_+  \\
 v_+  \\
\end{pmatrix} + A
\begin{pmatrix}
   \pm 1  \\
   1  \\
\end{pmatrix},
\end{gather}
where $A(t)$ denotes the amplitude of the dominating mode and the upper (lower) sign should be used if this mode is even (odd). The scattering matrix in the frequency domain is
\begin{equation}
\label{}
\mathcal{S}_{uv} = \frac {1} {1 - i (\delta \omega_{\mathrm{eff}} / \gamma)}
\begin{bmatrix}
   \pm i ( \delta \omega_{\mathrm{eff}} / \gamma ) & 1  \\
   1 & \pm i ( \delta \omega_{\mathrm{eff}} / \gamma )  \\
\end{bmatrix},
\end{equation}
where the effective frequency detuning was introduced $
\delta \omega_{\mathrm{eff}} = \omega - \omega_0 + \gamma |A|^2 /I_0
$
which takes into account the shift of the resonant frequency due to the influence of the Kerr nonlinearity. The amplitude $A$ depends on the amplitudes of the ingoing waves and can be found as a solution of the following equation
\begin{equation}
\label{eCubic}
\left[
1 - i \left(
\frac {\omega - \omega_0} {\gamma} + \frac {|A|^2} {I_0}
\right)
\right] A =
u_+ \pm v_+ .
\end{equation}
It is a cubic equation which has three different roots in general case. This agrees with the fact that the system can show several stable states for the same input signals and explains its bistable behavior from the mathematical point of view. It is also possible to find $A$ by using Eq.~(\ref{eCMTmain2}), which does not involve any nonlinear equations, however $A$ becomes a function of both ingoing and outgoing signals. This can be particularly suitable when signals are incident only from one side. For example, if the incidence from the left is considered, $v_+ = 0$ and $A = -v_-$. Since the transmitted and input powers are given by $P_{\mathrm{out}} = \sigma |v_-|^2$ and $P_{\mathrm{in}} = \sigma |u_+|^2$, the nonlinear transmission spectrum in the vicinity of the resonance can be found as
\begin{equation}
\label{eTransmissionNonlinear}
T(\omega) = \frac {P_{\mathrm{out}}} {P_{\mathrm{in}}} =
\left[1 + \left(
\frac {\omega - \omega_0} {\gamma} +
\frac {P_{\mathrm{out}}} {P_0}
\right)^2 \right]^{-1}.
\end{equation}
For a fixed value of the frequency detuning, the formula (\ref{eTransmissionNonlinear}) can be used to compute the hysteresis curve. It can be checked that this curve is equivalent to the polynomial of the third degree with real coefficients, and thus it can describe only bistable resonances. More complex structures which demonstrate multistable behavior can be constructed by combining several strictly bistable microcavities \cite{Grigoriev2010}. To treat the nonlinear properties of such structures, it is useful to define the nonlinear transfer matrix $\vv{T}_{uv}$
\begin{gather}
\label{eTmatrixDefinition}
\begin{pmatrix}
   u_+  \\
   u_-  \\
\end{pmatrix} = \vv{T}_{uv}
\begin{pmatrix}
   v_-  \\
   v_+  \\
\end{pmatrix}, \\
\label{eTmatrixResult}
\vv{T}_{uv} = \vv{I} - i \left(
\frac {\omega - \omega_0} {\gamma} +
\frac {|v_+ \pm v_-|^2} {I_0}
\right)
\left[ \begin{array}{cc}
   1 & \pm 1  \\
   \mp 1 & -1  \\
\end{array} \right],
\end{gather}
where $\vv{I}$ is the identity matrix. The transfer matrices of single microcavities can be multiplied producing the total transfer matrix of the structure. This leads to a hysteresis curve of more complicated shape which can be different for the opposite directions of incidence because the transfer matrices do not commute and the order of multiplication plays an important role \cite{Grigoriev2011a}.

\section{Numerical examples\label{sExamples}}
\subsection{Bragg gratings with symmetrically placed defect\label{sExamples`sBragg}}
Multilayered structures can be considered as a one-dimensional (1D) realization of on-channel microcavities and are particularly suitable to check the accuracy of the CMT equations. As a test case, we use a Bragg structure with a symmetrically placed defect which can be described by the symbolic formula $(\mathrm{HL})^p (\mathrm{LH})^p$. In what follows, the letters '$\rm{L}$' and '$\rm{H}$' correspond to quarter-wave layers of polydiacetylene 9-BCMU with linear (nonlinear) refractive index  $n_{\rm{L}} = 1.55$ ($n_{2\rm{L}} = 2.5 \times 10^{-5} \; \rm{cm^2/MW}$) and rutile with $n_{\rm{H}} = 2.3$ ($n_{2\rm{H}} = 10^{-8} \; \rm{cm^2/MW}$), respectively~\cite{Tocci1995, Biancalana2008}. The quarter wave condition is set to $\lambda_{\mathrm{q}} = 0.7 \; \rm{\mu m}$
\begin{equation}
\label{eQWcondition}
n_{\mathrm{L}} d_{\mathrm{L}} =
n_{\mathrm{H}} d_{\mathrm{H}} =
\lambda_{\mathrm{q}} / 4,
\end{equation}
which gives the thicknesses of the layers $d_{\rm{L}} = 112 \; \rm{nm}$ and $d_{\rm{H}} = 76 \; \rm{nm}$. The main advantage of this structure is that it has a well-defined resonance at $\omega_{\mathrm{q}} = 2 \pi c / \lambda_{\mathrm{q}}$ which is surrounded by band gap regions. Due to the mirror symmetry of the structure, this resonance always shows perfect transmission, and its half-width can be adjusted by the number of periods $p$ in the Bragg mirrors.

The formulas for the decay rate (\ref{eGamma3D}) and the characteristic intensity (\ref{eCharacteristicPowers}) of the microcavity at the resonance $\lambda_0 = 2 \pi c / \omega_0$ can be simplified in the 1D case to
\begin{gather}
\label{eGamma1D}
\gamma = \left[
(1 / c) {\varint}_{\!0}^L
\varepsilon E_0^2 \mathrm{d}x
\right]^{-1}, \\
\label{eCharacteristicIntensity1D}
I_0 = \left[
(2 \pi/ \lambda_0) {\varint}_{\!0}^L \varepsilon n_2 E_0^4 \mathrm{d}x
\right]^{-1}.
\end{gather}
Instead of the decay rate $\gamma$, it is often convenient to use a dimensionless quality factor defined as $Q = \omega_0 / (2 \gamma)$
\begin{equation}
\label{eQfactor}
Q = (\pi / \lambda_0)
{\varint}_{\!0}^L \varepsilon E_0^2 \mathrm{d}x.
\end{equation}
In some simple cases, the integration can be performed analytically, and an explicit formula for the quality factor can be obtained [see Appendix
].

For the structure $(\mathrm{HL})^8 (\mathrm{LH})^8$, the quality factor computed by Eq.~(\ref{eQfactor}) is $Q = 2060$ and the characteristic intensity according to Eq.~(\ref{eCharacteristicIntensity1D}) is  $I_0 = 0.05695 \; \mathrm{MW} / \mathrm{cm}^2$. The two parameters together with Eq.~(\ref{eTransmissionNonlinear}) fully determine the hysteresis and the nonlinear transmission spectrum of the structure in the vicinity of the resonant frequency $\omega_0 = \omega_{\mathrm{q}}$. The quality factor can be also computed with the linear transfer matrix by finding the full-width at half-maximum of the resonance $\omega_0 / \Delta \omega_{\mathrm{FWHM}} = 2058$. Therefore, the accuracy of the CMT equations in this case can be estimated as 0.1\%.

\begin{figure}
\center{\includegraphics[width=80mm]{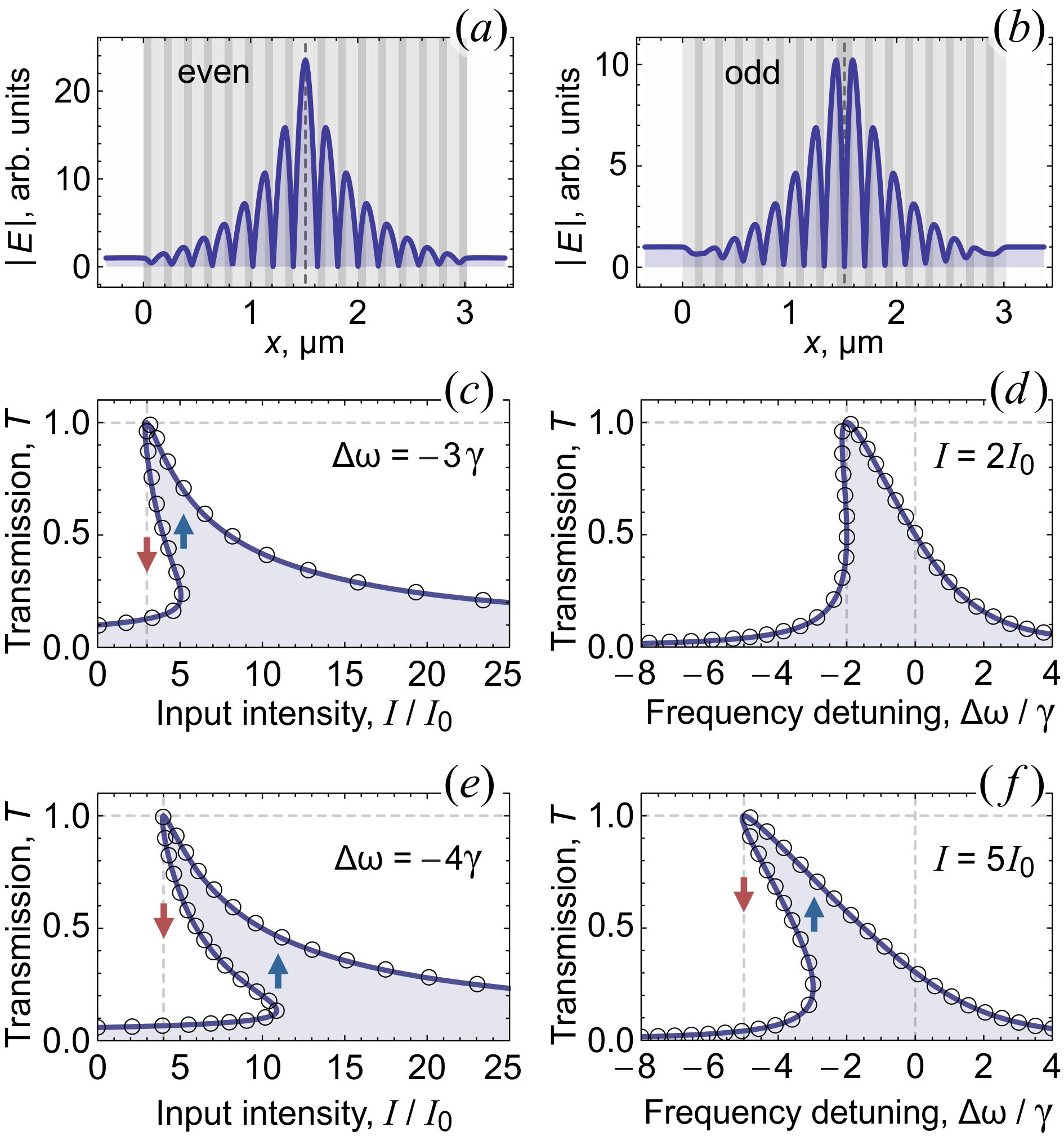}}
\caption{\label{figBragg}(Color online).
The structures $(\textrm{HL})^8(\textrm{LH})^8$ (a) and $(\textrm{LH})^8(\textrm{HL})^8$ (b) show a strong resonance at the quarter wave frequency  $\omega_0 = \omega_{\mathrm{q}}$, but have a different distribution of the electric field.
The hysteresis of transmission for a fixed value of the frequency detuning (c,~e) can be computed by Eq.~(\ref{eTransmissionNonlinear}), which follows from the CMT (solid lines), and by solving the Maxwell equations (\ref{eRungeKutta1},~\ref{eRungeKutta2}) directly with the Runge-Kutta method (circles). Not all parts of the hysteresis curves are stable, and arrows show how the switching between different branches occurs. The usage of normalized units ensures that both structures have the same shape of the hysteresis curve. The results of the two methods are in good agreement (the comparison for $(\textrm{HL})^8(\textrm{LH})^8$ is shown, $(\textrm{LH})^8(\textrm{HL})^8$ is similar). The same methods can be applied to compute the nonlinear transmission spectrum for a fixed value of the input intensity (d,~f).
}
\end{figure}

It is worth noting that the CMT parameters of a similar structure $(\mathrm{LH})^8 (\mathrm{HL})^8$ are different. It has a smaller quality factor $Q = 581.3$, and a significantly larger characteristic intensity $I_0 = 1.593 \; \mathrm{MW} / \mathrm{cm}^2$. This can be explained by a different localization of the electric field in the structure [Fig.~\ref{figBragg}(a,b)]. Nevertheless, the usage of normalized units ensures that both structures have the same shape of the hysteresis and the nonlinear transmission spectrum [Fig.~\ref{figBragg}(c--f)].

The Maxwell equations for nonlinear multilayered structures
\begin{align}
\label{eRungeKutta1}
\partial_x E_y &= i (\omega /c) H_z, \\
\label{eRungeKutta2}
\partial_x H_z &= i (\omega /c) \varepsilon (1 + 2n_2 |E_y|^2) E_y
\end{align}
can be also solved by using strictly numerical methods \cite{Press2007, Trutschel1989} or semi-analytical techniques \cite{Chen1987, Gupta1987}. A detailed comparison with the results obtained by the Runge-Kutta method is presented in Fig.~\ref{figBragg}(c--f).

\subsection{Thue-Morse multilayered structures\label{sExamples`sThue-Morse}}
\begin{figure}
\center{\includegraphics[width=80mm]{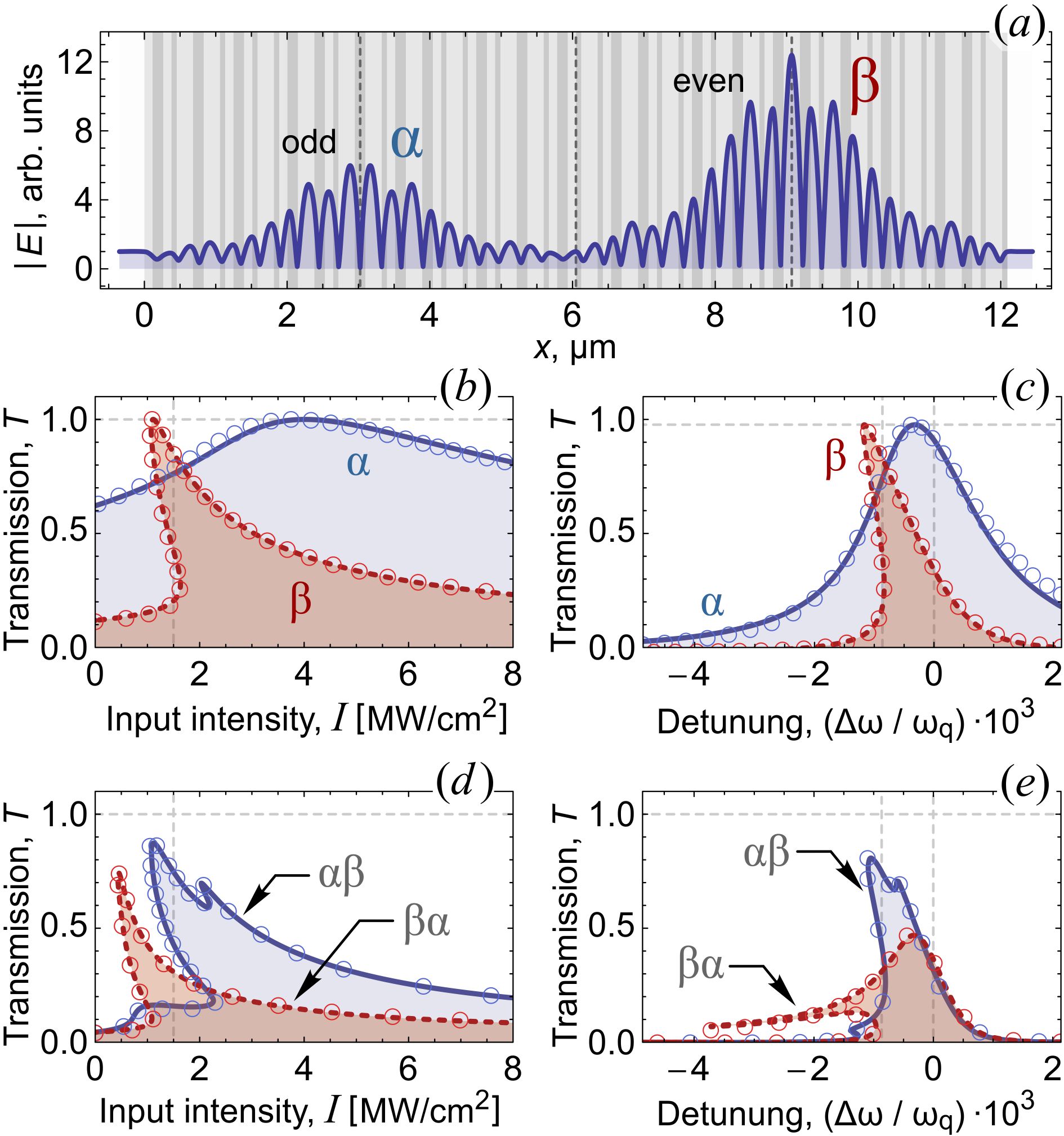}}
\caption{\label{figThue-Morse}
(Color online).
An example of the resonance (the Thue-Morse structure of the 7th generation number at the frequency $\omega_0 = 0.705465 \; \omega_{\mathrm{q}}$) which has two localization centers in the electric field profile (a) denoted as $\alpha$ and $\beta$. The hysteresis (b) and nonlinear transmission spectrum (c) of these parts can be computed by Eq.~(\ref{eTmatrixResult}), which follows from the CMT (solid lines), and by solving the Maxwell equations (\ref{eRungeKutta1},~\ref{eRungeKutta2}) directly with the Runge-Kutta method (circles). The discrepancy between the two methods is noticeable only for the part $\alpha$, because it has a relatively small localization strength. The Thue-Morse structure, which combines the parts $\alpha$ and $\beta$ together, shows the nonreciprocal behavior. For the same frequency $\omega = 0.7046 \; \omega_{\mathrm{q}}$ and input intensity $I = 1.5 \; \mathrm{MW} / \mathrm{cm}^2$ as used in (b) and (c), respectively, the hysteresis (c) and nonlinear transmission spectrum (e) depend on the propagation direction ($\alpha\beta$ or $\beta\alpha$).
}
\end{figure}

As a more complex example, we consider a Thue-Morse quasicrystal. It has a nonperiodic arrangement of layers which is governed by a deterministic set of inflation rules and features a number of pseudo band gap regions with resonances of complete transmission \cite{Grigoriev2010}. We choose one of such resonances which is located at $\omega_0 = 0.705465 \; \omega_{\mathrm{q}}$. It has two localization centers in the field profile so that the full structure can be divided into two parts which can be treated as coupled microcavities [Fig.~\ref{figThue-Morse}(a)]. These microcavities, which will be denoted as $\alpha$ and $\beta$, have the same resonant frequencies $\omega_{\alpha} = \omega_{\beta} = \omega_0$, but their parity and CMT parameters are different. For the left part, the quality factor is $Q_\alpha = 318.2$ and the characteristic intensity is $I_\alpha = 5.147 \; \mathrm{MW} / \mathrm{cm}^2$, while for the right part $Q_\beta = 1110$ and $I_\beta = 0.4078 \; \mathrm{MW} / \mathrm{cm}^2$.

The nonlinear response of these microcavities is qualitatively similar and can be described by the same analytic formula (71) [Fig.~\ref{figThue-Morse}(b,c)]. The discrepancy with the numerical results is noticeable only for the microcavity $\alpha$ which has a relatively small quality factor. Since the accuracy is worse on the higher frequency side of the resonance where the band gap is less pronounced, this suggests that the influence of other resonances causes additional perturbations.

The coupling between microcavities leads to more complex hysteresis curves and nonlinear transmission spectra [Fig.~\ref{figThue-Morse}(d,e)]. They can be obtained by multiplying the nonlinear transfer matrices of the microcavities (\ref{eTmatrixResult}) for a fixed output intensity and then restoring the input intensity. The order of multiplication plays an important role in the nonlinear case because the transfer matrices do not commute and the result strongly depends on the direction of incidence. Apart from the nonreciprocal behavior, there is also the possibility of multistable behavior since the hysteresis curve in the case of coupled microcavities is described by a polynomial of a higher degree. It is very important that a simple model based on the CMT equations is able not only to explain the nonlinear properties of complex resonances like this one, but also shows a good quantitative agreement with computationally intensive numerical methods.

\section{Conclusions\label{sConclusions}}

The phenomenological CMT is a very efficient tool for studying the nonlinear behavior of microcavities both in the frequency and time domain. It considers the interaction between microcavity and waveguide modes in a way that is similar to the scattering formalism. Therefore, the complex wave dynamics can be separated from a relatively simple picture of coupling, and this gives a significant advantage in comparison to strictly numerical methods. The dynamical properties of the microcavities can be fully determined by a small set of parameters which includes the decay rate, coupling coefficients and characteristic intensities.

By using on-channel microcavities with two coupling ports as an example, we provided for the first time a systematic derivation of the CMT equations starting directly from the Maxwell equations and obtained the explicit formulas for all phenomenological parameters. Our derivation is particularly suitable for microcavities embedded in photonic crystal waveguides of various dimensionality and multilayered structures. The accuracy of the results depends on the quality factor of a specific resonance and is mostly limited by the influence of other resonances.

\ifx \myStyle \styleJOSAB
\section*{Appendix A: Quality factors of Bragg gratings with symmetrically placed defects\label{sQfactors}}
\fi
\ifx \myStyle \stylePRA
\appendix*
\section{Quality factors of Bragg gratings with symmetrically placed defects\label{sQfactors}}
\fi
By using the fact that the energy density is a constant in each layer of the structure, the formula for the quality factor (\ref{eQfactor}) can be rewritten as
\begin{equation}
\label{eQfactor2}
Q = \frac {\pi} {2 \lambda_0}
\sum\limits_k
{(n_k^2 |E_k|^2 + |H_k|^2) d_k},
\end{equation}
where $d_k$ is the thickness of the layer $k$ with the refractive index $n_k$, and the sum is taken over all layers in the structure. Fields on opposite sides of the layer $k$ can be related by the characteristic matrix $\vv{M}_k$
\begin{gather}
\label{}
\left( \begin{array}{c}
   E_{k - 1}  \\
   H_{k - 1}  \\
\end{array} \right) = {\vv{M}}_k
\left( \begin{array}{c}
   E_k  \\
   H_k  \\
\end{array} \right), \\
\label{}
{\vv{M}}_k  = \left[ \begin{array}{cc}
   \cos \xi_k  &  i n_k^{-1} \sin \xi_k  \\
   i n_k \sin \xi_k  &  \cos \xi_k   \\
\end{array} \right],
\end{gather}
where $\xi_k = n_k d_k \omega /c$, or $\xi_k  = (\pi \omega ) / (2 \omega_{\mathrm{q}})$ if all layers satisfy the quarter wave condition (\ref{eQWcondition}) at the frequency $\omega_{\mathrm{q}}$.
The M-matrix for the single period of the Bragg gratings in the structure $(\mathrm{HL})^p (\mathrm{LH})^p$ can be obtained as a multiplication of M-matrices corresponding to layers 'L' and 'H'. It takes a particularly simple form at the resonance $\omega_0 = \omega_{\mathrm{q}}$
\begin{equation}
\label{}
{\vv{M}}_{\mathrm{L}} {\vv{M}}_{\mathrm{H}} = - \left[ \begin{array}{cc}
   n_{\mathrm{H}} / n_{\mathrm{L}}  & 0  \\
   0 & n_{\mathrm{L}} / n_{\mathrm{H}}  \\
\end{array} \right],
\end{equation}
which shows that the fields are exponentially growing or decaying towards the center of the structure as $(n_{\mathrm{H}} / n_{\mathrm{L}})^p$, where $p$ is the number of periods in the Bragg mirrors. The contribution of each period to the quality factor is
\begin{equation}
\label{}
\frac {\pi} {8}
\left[
\left( n_{\mathrm{H}} + \frac {n_{\mathrm{H}}^2} {n_{\mathrm{L}}} \right) |E_k|^2 +
\left( \frac {1} {n_{\mathrm{H}}} + \frac {n_{\mathrm{L}}} {n_{\mathrm{H}}^2} \right) |H_k|^2
\right],
\end{equation}
which makes in total
\begin{equation}
\label{}
Q = \frac {\pi} {4} (n_{\mathrm{H}} + n_{\mathrm{L}})
{\sum\limits_{k = 0}^{p - 1}} {\left[
\left(
\frac {n_{\mathrm{H}}} {n_{\mathrm{L}}}
\right)^{2k+1}\!\!\!\!\!\! +
\frac {1} {n_{\mathrm{H}}^2}
\left(
\frac {n_{\mathrm{L}}} {n_{\mathrm{H}}}
\right)^{2k}
\right]}.
\end{equation}
The sum of the geometric progressions can be found as $
\sum\nolimits_{k = 0}^{p - 1} r^k = (1 - r^p ) / (1 - r)
$ and keeping only the largest term leads to the following formula for the quality factor of the structure $(\mathrm{HL})^p (\mathrm{LH})^p$
\begin{equation}
\label{}
Q = \frac {\pi n_{\mathrm{H}} n_{\mathrm{L}}} {4 (n_{\mathrm{H}} - n_{\mathrm{L}})}
\left( \frac {n_{\mathrm{H}}} {n_{\mathrm{L}}} \right)^{2p}.
\end{equation}
It is worth noting that the quality factor of a similar structure $(\mathrm{LH})^p (\mathrm{HL})^p$
\begin{equation}
\label{}
Q = \frac {\pi} {4 (n_{\mathrm{H}} - n_{\mathrm{L}})}
\left( \frac {n_{\mathrm{H}}} {n_{\mathrm{L}}} \right)^{2p}
\end{equation}
is smaller in $n_{\mathrm{H}} n_{\mathrm{L}}$ times.

\ifx \myStyle \styleJOSAB
\section*{Acknowledgments}
This work was supported by the German Max Planck Society for the Advancement of Science (MPG).
\fi

\ifx \myStyle \stylePRA
\begin{acknowledgments}
This work was supported by the German Max Planck Society for the Advancement of Science (MPG).
\end{acknowledgments}
\fi

\ifx \myStyle \styleJOSAB

\fi

\ifx \myStyle \stylePRA
\bibliography{references}
\fi

\end{document}